\documentclass{elsart}
\usepackage{natbib,amsmath,amssymb,euscript}
\usepackage[latin1]{inputenc}
\usepackage{gastex}
\usepackage{url}

\begin{document}

\newcommand{\Nat}{\mathbb{N}}
\newcommand{\Q}{\mathbb{Q}}
\newcommand{\Z}{\mathbb{Z}}
\newcommand{\Real}{\mathbb{R}}
\newcommand{\finiteset}[2]{\{#1\ldots#2\}}
\newcommand{\automaton}{A}
\newcommand{\Lan}{L}
\newcommand{\conv}{\mathrm{conv}}
\newcommand{\partie}{\mathcal{P}}
\newcommand{\starh}{\mathrm{starH}}
\newcommand{\crochet}[1]{[#1]}
\newcommand{\parikh}{\textrm{parikh}}

\newcommand{\trex}{\textsc{TreX}}
\newcommand{\fast}{\textsc{Fast}}
\newcommand{\lash}{\textsc{Lash}}
\newcommand{\babylon}{\textsc{Babylon}}
\newcommand{\brain}{\textsc{Brain}}
\newcommand{\cslalv}{\textsc{CSL-ALV}}
\newcommand{\mona}{\textsc{Mona}}
\newcommand{\fmona}{\textsc{FMona}}
\newcommand{\omegatool}{\textsc{Omega}}
\newcommand{\lcs}{\textsc{Lcs}}
\newcommand{\hytech}{\textsc{Hytech}}
\newcommand{\dif}{\backslash}

\newcommand{\scalar}[2]{\left<#1,#2\right>}
\newcommand{\norm}[2]{|#2|_{#1}}

\newcommand{\closure}{\mathrm{cl}}

\title{The convex hull of a regular set of integer vectors is polyhedral and effectively computable}

\begin{frontmatter}
  \author{Alain Finkel}
  \address{LSV, CNRS UMR 8643, ENS de Cachan, Cachan, France}
  %%Tel:(+33) 147 40 75 69\\
  %%Fax:(+33) 147 40 75 29
  \ead{finkel@lsv.ens-cachan.fr}
  
  \author{Jérôme Leroux}
  \address{ DIRO, Université de Montréal, Montréal, QC, Canada}
  \ead{leroujer@iro.umontreal.ca}
  \thanks{Research funded by the Faculté des arts et des sciences of the Université de Montréal and by the Natural Sciences and Engineering Research Council of Canada through a discovery grant held by Pierre McKenzie.}
  \begin{abstract}
    Number Decision Diagrams (NDD) provide a natural finite symbolic representation for regular set of integer vectors encoded as strings of digit vectors (least or most significant digit first). The convex hull of the set of vectors represented by a NDD is proved to be an effectively computable convex polyhedron.
\end{abstract}

\end{frontmatter}
Presburger arithmetic \cite{P-PCM29} is a decidable logic used in a large range of applications. Different techniques \cite{GBD-FMCAD02} and tools have been developed for manipulating \emph{the Presburger-definable sets} (the sets of integer vectors satisfying a Presburger formula): by working directly on the Presburger-formulas (implemented in \omegatool\ \cite{OMEGA}), by using semi-linear sets \cite{GS-PACIF66} (implemented in \brain\ \cite{RV-AMAST02}), or by using NDD (automata that represent regular sets of integer vectors encoded as strings of digit vectors, least or most significant digit first) \cite{BB-THESE98,WB-SAS95,BC-CAAP96} (implemented in \fast\ \cite{BFLP-CAV03}, \lash\ \cite{LASH} and \cslalv \cite{BB-IJFCS03}). Presburger-formulas and semi-linear sets lack canonicity: there does not exist a natural way to canonically represent a set. As a direct consequence, a set that possesses a simple representation could unfortunately be represented in an unduly complicated way. Moreover, deciding if a given vector of integers is in a given set, is at least \emph{NP-hard} \cite{B-FOCS77,GS-PACIF66}. On the other hand, a minimization procedure for automata provides a canonical representation for \emph{NDD-definable sets} (a set represented by a NDD). That means, the NDD that represents a given set only depends on the set and not on the way we have computed it. For this reason, NDD are well adapted for applications that require a lot of Boolean manipulations like model-checking.

Verification of systems with unbounded integer variables is undecidable in general. That explains why we are interested in over-approximating the reachability set of such a system. By computing the convex hull of the set of initial states of such a system and by using a widening operator \cite{cousot78,polka:fmsd:97}, an over-approximation of the set of reachability set can be effectively computed.

In this presentation, the convex hull of a set of integer vectors represented by a NDD is proved to be a convex polyhedron. That shows that it can be finitely represented as a \emph{finite intersection of half-spaces} or dually as a \emph{finite set of rays}. Indeed, we provide an \emph{exponential time algorithm} that effectively computes this convex hull (the exact complexity remains open).

This result is obtained by first proving that ``the convex hull'' of the language $\sigma_1^*.\sigma_2^*$ is equal to the convex hull of $\sigma_2^*.\sigma_1^*$ for any pair of words $(\sigma_1,\sigma_2)$. From this commutativity result, we deduce that the convex hull of any regular language $\Lan$, is equal to the convex hull of a finite union of regular languages of the form $w_0.\sigma_1^*\ldots w_{n-1}.\sigma_n^*.w_n$.

\section{Closed sets and convex sets}
Recall that the \emph{scalar product} of two real vectors $x,y\in\Real^m$ where $m\geq 1$ is the real $\scalar{x}{y}=\sum_{i=1}^mx[i].y[i]$ where $x[i]\in\Real$ corresponds to the $i$th component of $x$. We denote by $\norm{2}{x}$ the \emph{norm} $\norm{2}{x}=\sqrt{\scalar{x}{x}}$. The \emph{open ball} centered in $x\in\Real^m$ with a radius $\epsilon>0$ is the subset $B_{x,\epsilon}=\{y\in\Real^m;\;\norm{2}{x-y}<\epsilon\}$. Recall that a subset $X\subseteq \Real^m$ is said \emph{open} if for any $x\in X$ there exists $\epsilon>0$ such that $B_{x,\epsilon}\subseteq X$. A \emph{closed set} $X$ is a subset of $\Real^m$ such that difference $\Real^m\dif X$ is open. Recall that any infinite or finite intersection of closed sets is closed and any subset $X$ is included into a minimal (for the inclusion) closed set, called the \emph{closure} of $X$. We denote by $\closure:\partie(\Real^m)\rightarrow\partie(\Real^m)$ the function such that $\closure(X)$ is the closure of $X$ for any $X\subseteq\Real^m$.

An \emph{half-space} $H$ is a subset of real vectors $\Real^m$ such that there exists $\alpha\in\Real^m$ and $c\in\Real$ satisfying $H=\{x\in\Real^m;\;\scalar{\alpha}{x}+c\# 0\}$ where $\#\in\{\geq,>\}$. Recall that such an half space $H$ is closed if $\#$ is equal to $\geq$ and it is open if $\#$ is equal to $>$.

We denote by $\Real_+$ and $\Real_-$ respectively the set of non-negative reals $\Real_+=\{x\in\Real;\;x\geq 0\}$ and the set of non-positive reals $\Real_-\{x\in\Real;\;x\leq 0\}$.

A \emph{convex set} is a \emph{finite or infinite} intersection of half-spaces. The \emph{convex hull} of a subset $X\subseteq \Real^m$ is the least (for the inclusion $\subseteq$) convex set that contains $X$. We denote by $\conv:\partie(\Real^m)\rightarrow\partie(\Real^m)$ the function such that $\conv(X)$ is the convex hull of $X$ for any $X\subseteq\Real^m$. Recall that a vector $y$ is in $\conv(X)$ if and only if there exists a finite sequence $(x_i)_{1\leq i\leq n}$ of $n\geq 1$ vectors in $X$ and a sequence $(t_{i})_{1\leq i\leq n}$ of $n$ reals in $\Real_+$ such that $\sum_{i=1}^nt_i=1$ and such that $y=\sum_{i=1}^nt_i.x_i$. Recall that the closure of a convex set remains a convex set.

A convex set $C$ is said \emph{polyhedral} if $C$ is equal to a finite intersection of \emph{closed} half-spaces (in particular, a polyhedral convex set is closed). Recall that any polyhedral convex set $P$ can be represented by a finite set of \emph{rays} $R\subseteq \Real^m\times\Real_+$ such that $P=P(R)=\{x\in\Real^m;\;(x,1)\in C(R)\}$ where $C(R)\subseteq\Real^m\times\Real_+$ is the \emph{polyhedral cone} defined by the following equality:
$$C(R)=\{\sum_{r\in R}t_r.r;\;t_r\in\Real_+\}$$
Recall that for any pair $(P_1,P_2)$ of polyhedral convex sets respectively represented by a pair of finite set of rays $(R_1,R_2)$, the convex set $\closure\circ\conv(P_1\cup P_2)$ is polyhedral and represented by the set of rays $R_1\cup R_2$.

\section{Regular sets of integer vectors}\label{dvasection}
Let us consider an integer $r\geq 2$ called the \emph{basis of the decomposition} and an integer $m\geq 1$ called the \emph{dimension of the represented vectors}. A \emph{digit vector} $b$ is an element of the finite alphabet $\Sigma_{r^m}=\finiteset{0}{r-1}^m$. The vector $\rho(\sigma)\in\Nat^m$ associated with a word $\sigma=b_1\ldots b_n$ of $n\geq 1$ digit vectors $b_i\in\Sigma_{r^m}$ is defined by $\rho(\sigma)=\sum_{i=1}^{n}r^{i-1}.b_i$. We naturally define $\rho(\epsilon)=(0,\ldots,0)$, also written $0$.

The set $X$ represented by a \emph{language} $\Lan\subseteq\Sigma_{r^m}^*$ is defined by $X=\rho(\Lan)=\{\rho(\sigma);\;\sigma\in\Lan\}$. If $\Lan$ is \emph{regular} (that means accepted by a \emph{finite automaton}), the set $X$ is naturally said \emph{regular}. Let us recall that regular sets of vectors can be efficiently manipulated with finite automata (see \cite{WB-ICALP00,BC-CAAP96}) and they correspond to the sets defined by a formula in the first order logic $\left<\Nat,+,\leq,V_r\right>$ where $V_r$ is the valuation function in base $r$ defined by $y=V_r(x)$ if and only if $y$ is the greatest power of $r$ that divides $x$ \cite{BHMV-BMS94}.

\begin{exmp}
  Consider the following automaton $\automaton_+$ with basis $r=2$ and dimension $m=3$ depicted below. Intuitively, this automaton represents the set of vectors $(x,y,z)\in\Nat^3$ such that $x+y=z$ where the state $q_i$ corresponds to the carry $i\in\{0,1\}$ of the addition.
  
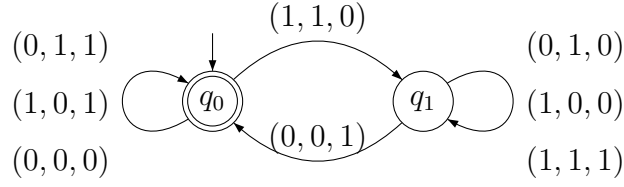
\begin{center}
    \vspace{-1cm}
    \begin{picture}(44,38)(0,-38)
      \node[NLangle=0.0,Nmarks=ri,iangle=90](n0)(8,-24){$q_0$}
      \node[NLangle=0.0](n1)(36,-24){$q_1$}
      \drawloop[loopangle=180](n0){%
        $\begin{array}{c}(0,1,1)\\(1,0,1)\\(0,0,0)\end{array}$}
      \drawloop[loopangle=0](n1){%
        $\begin{array}{c}(0,1,0)\\(1,0,0)\\(1,1,1)\end{array}$}
      \drawedge[curvedepth=8.0](n0,n1){$(1,1,0)$}
      \drawedge[ELside=r,curvedepth=8.0](n1,n0){$(0,0,1)$}
    \end{picture}
    \vspace{-1cm}
  \end{center}
\end{exmp}

\section{The convex hull of a regular set of integer vectors}\label{mainsection}
The main result of this paper is proved in this section. We show that the closure of the convex hull of a regular set of integer vectors is polyhedral and represented by a set of rays effectively computable in exponential time from any regular expression that defines this regular set.

As $\rho(\sigma w)=r^{|\sigma|}.\rho(w)+\rho(\sigma)$ for any pair of words $(\sigma,w)$, we introduce the function $\Gamma_\sigma:\Real^m\rightarrow\Real^m$ defined by $\Gamma_\sigma(x)=r^{|\sigma|}.x+\rho(\sigma)$. Remark that any regular language $\Lan$ can be decomposed into a finite union of regular languages of the form $\sigma_{n+1}.\Lan_{n}^*.\sigma_n\ldots\Lan_1^*.\sigma_{1}$ where $\sigma_i\in\Sigma_{r^m}^*$ and $\Lan_i\subseteq\Sigma_{r^m}^*$. The following lemma is a first step toward the computation of $\closure\circ\conv\circ\rho(\Lan)$.

\begin{prop}\label{prop:etoileconv}
  Let us consider a language $\Lan=\sigma_{n+1}.\Lan_{n}^*.\sigma_n\ldots\Lan_1^*.\sigma_{1}$ where $n\geq 0$,  $\sigma_i\in\Sigma_{r^m}^*$ and $\Lan_i\subseteq\Sigma_{r^m}^*$. We have the following equality:
  $$\closure\circ\conv\circ\rho(\Lan)=\Gamma_{\sigma_{n+1}\ldots\sigma_1}\circ\closure\circ\left(\{(0,\ldots,0)\}\cup\bigcup_{i=1}^n\Real_-.\Gamma_{\sigma_i\ldots\sigma_1}^{-1}\circ\closure\circ\conv\circ\xi(\Lan_i^*)\right)$$
  where $\sigma\rightarrow\xi(\sigma)$ is partially defined over $\Sigma_{r^m}^+$, by the following equality:
  $$\xi(\sigma)=\frac{\rho(\sigma)}{1-r^{|\sigma|}}$$
\end{prop}
\begin{pf}
  We denote by $C(\sigma_{n+1},\Lan_{n},\sigma_n,\ldots,\Lan_1,\sigma_1)$, the following set:
  $$\Gamma_{\sigma_{n+1}\ldots\sigma_1}\circ\closure\circ\left(\{(0,\ldots,0)\}\cup\bigcup_{i=1}^n\Real_-.\Gamma_{\sigma_i\ldots\sigma_1}^{-1}\circ\closure\circ\conv\circ\xi(\Lan_i^*)\right)$$
  Let us first prove inclusion (\ref{equ1}):
  \begin{equation}\label{equ1}
    C(\sigma_{n+1},\Lan_{n},\sigma_n,\ldots,\Lan_1,\sigma_1)\subseteq \closure\circ\conv\circ\rho(\sigma_{n+1}.\Lan_{n}^*.\sigma_n\ldots\Lan_1^*.\sigma_{1})
  \end{equation}
  If $n=0$, inclusion is immediate. Assume that $n\geq 1$ and let $i\in\finiteset{1}{n}$, we have just to show that $\Real_-.\Gamma_{\sigma_i\ldots\sigma_1}\circ\closure\circ\conv\circ\xi(\Lan_i^*)\subseteq\closure\circ\conv\circ\rho(\Lan)$. Naturally, if $\Lan_i\dif\{\epsilon\}=\emptyset$, this inclusion is immediate. Otherwise, let $w\in\Lan_i^*\dif\{\epsilon\}$. For any $k\in\Nat$, we have $\sigma_{n+1}\ldots\sigma_{i+1}.w^k.\sigma_i\ldots\sigma_1\subseteq\Lan$. From the following equality, we get $\Real_-.\Gamma_{\sigma_i\ldots\sigma_1}\circ\closure\circ\conv\circ\xi(\Lan_i^*)\subseteq\closure\circ\conv\circ\rho(\Lan)$:
  \begin{align*}
    \rho(\sigma_{n+1}\ldots\sigma_{i+1}.w^k.\sigma_i\ldots\sigma_1)
    &=\Gamma_{\sigma_{n+1}\ldots\sigma_1}((1-r^{k.|w|}).\Gamma_{\sigma_i\ldots\sigma_1}^{-1}(\xi(w))
  \end{align*}
  In particular, we have proved inclusion (\ref{equ1}). Let us prove the converse inclusion. Consider a sequence $(w_i)_{1\leq i\leq n+1}$ such that $w_i\in\Lan_i^*\dif\{\epsilon\}$. An immediate induction over $n\geq 0$, proves the following equality:
  $$\rho(\sigma_{n+2}.w_{n+1}.\sigma_{n+1}\ldots w_1.\sigma_1)=\Gamma_{\sigma_{n+1}\ldots\sigma_1}\left(\sum_{i=1}^{n+1}r^{|w_{n+1}\ldots w_{i+1}|}.(1-r^{|w_i|}).\Gamma_{\sigma_i\ldots\sigma_1}^{-1}\circ\xi(w_i)\right)$$
  As $r^{|w_{n+1}\ldots w_{i+1}|}.(1-r^{|w_i|})\in\Real_-$, we deduce the following inclusion:
  $$\closure\circ\conv\circ\rho(\sigma_{n+1}.(\Lan_{n}^*\dif\{\epsilon\}).\sigma_n\ldots(\Lan_1^*\dif\{\epsilon\}).\sigma_{1})\subseteq C(\sigma_{n+1},\Lan_{n},\sigma_n,\ldots,\Lan_1,\sigma_1)$$
  Naturally, from the previous inclusions taken over $n\geq 0$, we deduce the converse inclusion of (\ref{equ1}).
  \qed
\end{pf}

The previous proposition explains why we are interested in computing $\closure\circ\conv\circ\xi(\Lan^*)$ where $\Lan$ is a regular language. In fact, we have the following lemma.
\begin{lem}\label{lem:xietoile}
  For any $\Lan\subseteq\Sigma_{r^m}^*$, we have $\conv\circ\xi(\Lan^*)=\conv\circ\xi(\Lan)$.
\end{lem}
\begin{pf}
  From $\Lan\subseteq\Lan^*$, we deduce the inclusion $\conv\circ\xi(\Lan)\subseteq \conv\circ\xi(\Lan^*)$. Let us prove the converse inclusion. Let $w\in\Lan^*\dif\{\epsilon\}$. There exists a sequence $\sigma_1$, ..., $\sigma_k$ of $k\geq 1$ words in $\Lan\dif\{\epsilon\}$ such that $w=\sigma_1\ldots\sigma_k$. An immediate induction over $k\geq 1$ proves the following equality:
  $$\xi(\sigma_1\ldots\sigma_k)=\sum_{i=1}^kr^{|\sigma_1\ldots\sigma_{i-1}|}\frac{r^{|\sigma_i|}-1}{r^{|\sigma_1\ldots\sigma_k|}-1}.\xi(\sigma_i)$$
  As $\sum_{i=1}^k r^{|\sigma_1\ldots\sigma_{i-1}|}\frac{r^{|\sigma_i|}-1}{r^{|\sigma_1\ldots\sigma_k|}-1}=1$ and $r^{|\sigma_1\ldots\sigma_{i-1}|}\frac{r^{|\sigma_i|}-1}{r^{|\sigma_1\ldots\sigma_k|}-1}\in\Real_-$, we deduce that $\xi(w)\in\conv\circ\xi(\Lan)$. We deduce $\xi(\Lan^*)\subseteq\conv\circ\xi(\Lan)$ and by minimality of the convex hull of $\xi(\Lan^*)$, we get $\conv\circ \xi(\Lan^*)\subseteq\conv\circ\xi(\Lan)$.
  \qed
\end{pf}

Once again, we use the fact that a regular language $\Lan$ can be decomposed into a finite union of languages of the form $\sigma_{n+1}.\Lan_n^*.\sigma_{n}.\cdots\Lan_1^*.\sigma_1$.
\begin{prop}\label{prop:fin}
  Let us consider a language $\Lan=\sigma_{n+1}.\Lan_{n}^*.\sigma_n\ldots\Lan_1^*.\sigma_{1}$ where $n\geq 0$,  $\sigma_i\in\Sigma_{r^m}^*$ and $\Lan_i\subseteq\Sigma_{r^m}^*$. We have the following equality:
  $$\closure\circ\conv\circ\xi(\Lan)=\closure\circ\conv\left(\xi(\{\sigma_{n+1}\ldots\sigma_{1}\})\bigcup_{i=1}^n\Gamma_{\sigma_{i}\ldots\sigma_1}^{-1}\circ\closure\circ\conv\circ\xi(\Lan_i)\right)$$
\end{prop}
\begin{pf}
  Let us consider a language $\Lan$ of the form $\Lan=\sigma_2.\Lan_1^*.\sigma_1$ where $\sigma_1,\sigma_2\in\Sigma_{r^m}^*$ and $\Lan_1\subseteq\Sigma_{r^m}^*$ and let us prove the proposition for $\Lan$. Remark that if $\Lan_1\dif\{\epsilon\}=\emptyset$ or if $\sigma_2.\sigma_1=\epsilon$, lemma \ref{lem:xietoile} proves the proposition. So, we can assume that $\Lan_1\dif\{\epsilon\}\not=\emptyset$ and $\sigma_2.\sigma_1\not=\epsilon$. Let $C=\closure\circ\conv(\xi(\{\sigma_2.\sigma_1\})\cup\Gamma_{\sigma_1}^{-1}\circ\closure\circ\conv\circ\xi(\Lan_1))$ and consider $\sigma\in\Lan_1\dif\{\epsilon\}$ and $k\in\Nat$. We have the following equality:
  $$\xi(\sigma_2.\sigma^k.\sigma_1)=\frac{r^{|\sigma_2.\sigma_1|}-1}{r^{|\sigma_2.\sigma_1|+k.|\sigma|}-1}.\xi(\sigma_2.\sigma_1)+\frac{r^{|\sigma_2.\sigma_1|}.(r^{k.|\sigma|}-1)}{r^{|\sigma_2.\sigma_1|+k.|\sigma|}-1}.\Gamma_{\sigma_1}^{-1}(\xi(\sigma))$$
  In particular we deduce that $\Gamma_{\sigma_1}^{-1}\circ\xi(\sigma)\in\closure\circ\xi(\sigma_2.\sigma^*.\sigma_1)$. From $\sigma_2.\sigma^*.\sigma_1\subseteq \Lan$, we get $\Gamma_{\sigma_1}^{-1}\circ\xi(\Lan_1)\subseteq\closure\circ\xi(\Lan)$. And by minimality of the closure and the convex hull, we get $\Gamma_{\sigma_1}^{-1}\circ\closure\circ\conv\circ\xi(\Lan_1)\subseteq\closure\circ\conv\circ\xi(\Lan)$. From $\sigma_2.\sigma_1\in\Lan$, we deduce that $\xi(\{\sigma_{2}.\sigma_{1}\})\subseteq \xi(\Lan)$. We obtain $C\subseteq\closure\circ\conv\circ\xi(\Lan)$. Let us prove the converse inclusion. Consider $\sigma\in\Lan$. There exists $w\in\Lan_1^*$ such that $\sigma=\sigma_2.w.\sigma_1$. If $w=\epsilon$ then $\xi(\sigma)=\xi(\sigma_1.\sigma_2)\in C$. Otherwise, lemma \ref{lem:xietoile} proves that $\xi(w)\in\conv\circ\xi(\Lan_1)$. As $\xi(\sigma)=\frac{r^{|\sigma_2.\sigma_1|}-1}{r^{|\sigma_2.\sigma_1|+|w|}-1}.\xi(\sigma_2.\sigma_1)+\frac{r^{|\sigma_2.\sigma_1|}.(r^{|w|}-1)}{r^{|\sigma_2.\sigma_1|+|w|}-1}.\Gamma_{\sigma_1}^{-1}(\xi(w))$, we deduce that $\xi(\sigma)\in C$. We deduce that the other inclusion $\closure\circ\conv\circ\xi(\Lan)\subseteq C$. Therefore, the proposition is proved for $\Lan$.
  
  Now, assume the proposition proved for an integer $n\geq 1$ and let us consider a language $\Lan=\sigma_{n+2}.\Lan_{n+1}^*.\sigma_{n+1}\ldots\Lan_1^*.\sigma_{1}$ where $\sigma_i\in\Sigma_{r^m}^*$ and $\Lan_i\subseteq\Sigma_{r^m}^*$ and let us prove the proposition for $\Lan$. Consider $w_{n+1}\in\Lan_{n+1}^*$. As the proposition is proved for $n$, we deduce the following equality:
  \begin{align*}
    &\closure\circ\conv\circ\xi(\sigma_{n+2}.w_{n+1}.\sigma_{n+1}.\Lan_{n}^*.\sigma_n\ldots \Lan_1^*.\sigma_1)\\
    &=\closure\circ\conv\left(\xi(\{\sigma_{n+2}.w_{n+1}.\sigma_{n+1}\ldots\sigma_{1}\})\bigcup_{i=1}^n\Gamma_{\sigma_{i}\ldots\sigma_1}^{-1}\circ\closure\circ\conv\circ\xi(\Lan_i)\right)
  \end{align*}
  We get in particular the following equality:
  \begin{align*}
    &\closure\circ\conv\circ\xi(\sigma_{n+2}.\Lan_{n+1}^*.\sigma_{n+1}.\Lan_{n}^*.\sigma_n\ldots \Lan_1^*.\sigma_1)\\
    &=\closure\circ\conv\left(\xi(\sigma_{n+2}.\Lan_{n+1}^*.\sigma_{n+1}\ldots\sigma_{1}\})\bigcup_{i=1}^n\Gamma_{\sigma_{i}\ldots\sigma_1}^{-1}\circ\closure\circ\conv\circ\xi(\Lan_i)\right)\\
    &=\closure\circ\conv\left(\closure\circ\conv\circ\xi(\sigma_{n+2}.\Lan_{n+1}^*.\sigma_{n+1}\ldots\sigma_{1})\bigcup_{i=1}^n\Gamma_{\sigma_{i}\ldots\sigma_1}^{-1}\circ\closure\circ\conv\circ\xi(\Lan_i)\right)
  \end{align*} 
  As the proposition is proved in the case $n=1$, we also get the following equality:
  \begin{align*}
    &\closure\circ\conv\circ\xi(\sigma_{n+2}.\Lan_{n+1}^*.\sigma_{n+1}\ldots\sigma_{1})\\
    &=\closure\circ\conv\left(\xi(\{\sigma_{n+2}\ldots\sigma_1\})\cup\Gamma_{\sigma_{n+1}\ldots\sigma_1}^{-1}\circ\closure\circ\conv\circ\xi(\Lan_{n+1})\right)
  \end{align*} 
  The two previous equality proved the proposition for $\Lan$. By induction over $n\geq 1$, we have proved the proposition for any $n\geq 1$.
  \qed
\end{pf}

We can now prove our main result that extends \cite{L-LICS04}.
\begin{thm}
  The convex hull of a regular set of integer vectors $X=\rho(\Lan)$ is polyhedral and a finite set of rays $R$ that represents $\conv(X)$ can be computed in exponential time from any regular expression that defines $\Lan$.
\end{thm}
\begin{pf}
  Let $C$ be a polyhedral convex set represented by a finite set of rays $R$. We know that for any $w\in\Sigma_{r^m}^*$, the convex sets $\Gamma_{w}(C)$ and $\Gamma_{w}^{-1}(C)$ are polyhedral and respectively represented by $\{(r^{|w|}.\alpha+c.\rho(w),c);\;(\alpha,c)\in R\}$ and $\{(\alpha-c.\rho(w),r^{|w|}.c);\;(\alpha,c)\in R\}$. Moreover, we also know that the convex set $\closure(\Real_-.C)$ is polyhedral and represented by the finite set of rays $\{(0,1)\}\cup\{(-\alpha,0);\;(\alpha,c)\in R\}$. By applying propositions \ref{prop:etoileconv} and \ref{prop:fin} and lemma \ref{lem:xietoile} over a regular expression that represents a regular language $\Lan$, we deduce that $\closure\circ\conv\circ\rho(\Lan)$ is polyhedral and represented by a finite set of rays $R$ computable in exponential time from any regular expression that defines $\Lan$.
  \qed
\end{pf}

\bibliography{biblio-these}
\bibliographystyle{alpha}

\end{document}